\documentclass[letterpaper,prl,superscriptaddress,showpacs,floats,twocolumn,ps2pdf]{revtex4}
\usepackage[latin1]{inputenc}
\usepackage[T1]{fontenc}

\usepackage{amsmath,amssymb}
\usepackage{graphicx,color}
\usepackage{hyperref}
\usepackage{hypernat}

\definecolor{rltred}{rgb}{0.75,0,0}
\definecolor{rltgreen}{rgb}{0,0.5,0}
\definecolor{rltblue}{rgb}{0,0,0.75}
\hypersetup{colorlinks,%
hypertexnames=true,%
pdfauthor={Johannes Feist, Stefan Nagele, Renate Pazourek, Emil Persson, Barry I. Schneider, Lee A. Collins, Joachim Burgd\"orfer},%
pdftitle={Probing Electron Correlation via Attosecond XUV Pulses in the Two-Photon Double Ionization of Helium},%
pdfview=FitH,%
pdfstartview=FitH,%
pdfpagemode=UseNone,%
bookmarksopen=true,%
bookmarksnumbered=true,%
pdfhighlight=/I,%
linkcolor=rltred,%
citecolor=rltgreen,%
linktocpage=true}

\newcommand{\Int}{\int\limits}

\newcommand{\ie}{i.e., }
\newcommand{\Schro}{Schr\"o\-din\-ger }

\newcommand{\cf}{cf.\@ }

\newcommand{\abs}[1]{\left|#1\right|}
\newcommand{\cvec}[1]{\mathbf{#1}}
\newcommand{\eqcomma}{\,,\;\;}

\newcommand{\dd}{\mathrm{d}}
\newcommand{\dW}{\dd\Omega}
\newcommand{\dE}{\dd E}
\newcommand{\hw}{\hbar\omega}

\newcommand{\ev}{\,\text{eV}}
\newcommand{\au}{\,\text{a.u.}}
\newcommand{\as}{\,\text{as}}
\newcommand{\fs}{\,\text{fs}}

\newcommand{\ti}{{t_\text{(i)}}}
\newcommand{\Tp}{{T_{\text{p}}}}

\begin{document}

\title{Probing Electron Correlation via Attosecond XUV Pulses in the Two-Photon Double Ionization of Helium}

\author{J.~Feist}
\email{johannes.feist@tuwien.ac.at} 
\affiliation{Institute for Theoretical Physics, 
             Vienna University of Technology, 1040 Vienna, Austria, EU}
\author{S.~Nagele}
\affiliation{Institute for Theoretical Physics, 
             Vienna University of Technology, 1040 Vienna, Austria, EU}
\author{R.~Pazourek}
\affiliation{Institute for Theoretical Physics, 
             Vienna University of Technology, 1040 Vienna, Austria, EU}
\author{E.~Persson}
\affiliation{Institute for Theoretical Physics, 
             Vienna University of Technology, 1040 Vienna, Austria, EU}
\author{B.~I.~Schneider}
\affiliation{Physics Division, 
             National Science Foundation, Arlington, Virginia 22230, USA}
\affiliation{Electron and Atomic Physics Division, 
             National Institute of Standards and Technology, Gaithersburg, Maryland 20899, USA}
\author{L.~A.~Collins}
\affiliation{Theoretical Division, T-4,
             Los Alamos National Laboratory, Los Alamos, New Mexico 87545, USA}
\author{J.~Burgd\"orfer}
\affiliation{Institute for Theoretical Physics, 
             Vienna University of Technology, 1040 Vienna, Austria, EU}

\date{\today}

\begin{abstract}
Recent experimental developments of high-intensity, short-pulse XUV
light sources are enhancing our ability to study
electron-electron correlations. We perform time-dependent calculations
to investigate the so-called ``sequential'' regime ($\hw\!>\!54.4\ev$) in
the two-photon double ionization of helium. We show that attosecond
pulses allow to induce and probe angular and energy correlations of the emitted
electrons. The final momentum distribution reveals regions dominated by the Wannier ridge break-up scenario
and by post-collision interaction.
\end{abstract}
\maketitle

Understanding the role of electron correlation in atoms, molecules, and
solids has been a central theme in physics and chemistry since the
early days of quantum mechanics. Most of the focus has
centered on the role of electron correlation in (quasi-)stationary states.
Recent progress in the development of light sources
provides unprecedented opportunities to expand
our understanding of electron correlation to 
dynamical processes where external fields play a critical role. 
The availability of atto\-second pulses, as generated from high-harmonic radiation
\cite{HenKieSpi2001,SanBenCal2006,GouSchHof2008}, opens new avenues for 
time-domain studies of multi-electron dynamics. Using such pulses, it is
possible to not only observe, but also actively \emph{induce} and \emph{control} correlation effects.

The simplest system where electron-electron interaction can be studied is the helium atom.
Unraveling the intricacies of electron correlation in ultrashort and intense 
electromagnetic fields interacting with this simple atom is critical to our understanding of the same
processes in more complex systems. 
Despite the computational challenges, the dynamics of He under 
the influence of external fields can still be accurately simulated in 
\emph{ab initio} calculations, \cf \cite{ParSmyTay1998}. 
The results of the present investigation provide
evidence that the effects of electron correlation can be surprisingly complex in
situations dominated by external ultrashort fields. This in turn has important consequences
for atto\-second studies in molecules, clusters, and solids. We show
that it is possible to disentangle the different processes occurring in 
such pulses by analyzing the final momentum distribution of the ejected 
electrons.

Double ionization of helium by single photon absorption has long
been the benchmark for our understanding of correlation effects in the
three-body Coulomb problem
\cite{ByrJoa1967,ProSha1993,BraDoeCoc1998,BriSch2000,MalSelKaz2000}. 
The availability of intense light sources in the VUV and XUV region
\cite{FLASH2007,NabHasTak2005,DroZepGop2006} 
has recently shifted attention from single-photon double ionization and intense-IR laser
ionization by rescattering (see \cite{LeiGroEng2000a,StaRuiSch2007,RudJesErg2007} and references therein) to 
multiphoton ionization. Restricting attention to only two-photon
double ionization (TPDI) enables us to distinguish two spectral regions. The
``nonsequential'' or ``direct'' regime between $39.5\ev < \hw < 54.4\ev$
has recently received considerable attention (see
\cite{NikLam2007,ProManMar2007,HorMccRes2008,FeiNagPaz2008,AntFouPir2008,GuaBarSch2008} and
references therein). Energy-sharing between the electrons, and thus
correlations, are a \emph{conditio sine qua non} for double ionization
to occur in this regime. By contrast, in ``sequential'' TPDI
with $\hw > 54.4\ev$ \cite{IshMid2005,BarWanBur2006,FouAntBac2008,PirBauLau2003},
each photon has sufficient energy to
ionize one electron within an independent-particle model and
electron-electron interaction, while present, is not a necessary
prerequisite.

For an ultrashort pulse of attosecond duration the
concept of ``sequential interactions'', valid for long pulses, becomes
obsolete. Instead, the two-electron emission occurs almost
simultaneously, and the strength of electron correlation in the exit
channel can be tuned by the pulse duration $\Tp$. This information is
encoded in the final joint momentum distribution
$P(\cvec{k}_1,\cvec{k}_2) \equiv P(E_1,E_2,\Omega_1,\Omega_2)$, experimentally
accessible in kinematically complete COLTRIMS measurements
\cite{UllMosDor2003}. 

In our current calculations we solve the time-dependent \Schro equation in its full
dimensionality, including all inter-particle interactions.
The laser field is linearly polarized and treated in dipole approximation. 
The duration $\Tp$ is given by the FWHM of a sine-squared envelope function for the electromagnetic field.
The computational approach is based on
a time-dependent close-coupling (TDCC) scheme where the angular variables
are expanded in coupled spherical harmonics and the two radial 
variables are discretized via a finite element discrete variable representation
(FEDVR). Temporal propagation is performed by the short
iterative Lanczos (SIL) algorithm with adaptive time-step
control. The asymptotic momentum distribution is obtained by projecting the
wave packet onto products of Coulomb continuum
states. Projection errors due to the replacement of the full
three-body final state by independent-particle Coulomb wave functions
can be reduced to the one-percent level by delaying the
time of projection until the two electrons are sufficiently far apart
from each other \cite{FeiNagPaz2008}.

\begin{figure}[tbp]
  \centering
  \resizebox{\linewidth}{!}{
    \includegraphics[height=8cm]{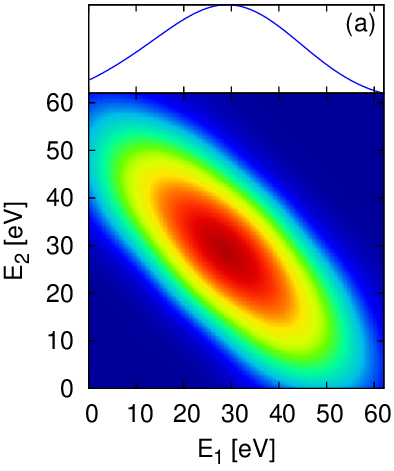}
    \includegraphics[height=8cm]{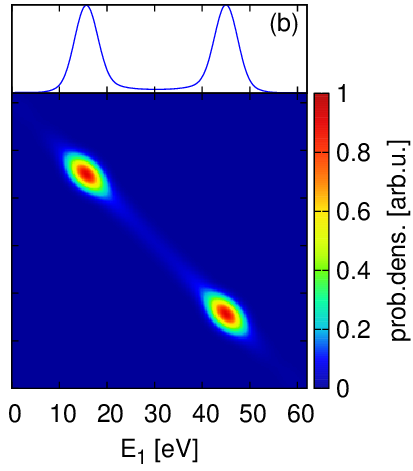}
  }
  \caption{TPDI electron spectra $P(E_1,E_2)$ at
  $\hw\!=\!70\ev$ for different pulse durations (FWHM): (a) $\Tp\!=\!150\as$,
  (b) $\Tp\!=\!750\as$. The top shows the
  spectrum integrated over one energy, \ie the one-electron energy
  spectrum $P(E_1)\!=\!P(E_2)$.}  \label{fig:energy_spectra}
\end{figure}

The joint energy probability distribution
\begin{equation}
P(E_1,E_2)=\iint P(E_1,E_2,\Omega_1,\Omega_2)\dW_1\dW_2
\end{equation}
reveals the breakdown of
the sequential ionization picture with decreasing pulse duration $\Tp$
(\autoref{fig:energy_spectra}). For long pulses, two
distinct peaks signifying the emission of the ``first'' electron with
energy $E_1=\hw-I_1$ (with $I_1\!=\!24.6\ev$ the first ionization potential) and
the ``second'' electron with $E_2=\hw-I_2$ ($I_2\!=\!54.4\ev$) are clearly visible.

For pulses of the order of
one hundred attoseconds a dramatically different picture emerges: the
two peaks merge into a single one located near the point of symmetric energy
sharing.
It should be noted that this is not simply due to the Fourier
broadening of the pulse. Instead, the close proximity
in time of the two emission events allows for energy exchange between
the two outgoing electrons representing a clear departure from the
independent-particle behavior \cite{IshMid2005,PirBauLau2003}. Differently stated, the time interval
between the two ionization events is too short for the ``remaining''
electron to relax to a stationary ionic state. In
the limit of ultrashort pulses the notion of a definite time ordering
of emission processes loses its significance, as does the distinction
between ``sequential'' and ``nonsequential'' ionization.

\begin{figure}[tb]
  \centering
  \includegraphics[width=\linewidth]{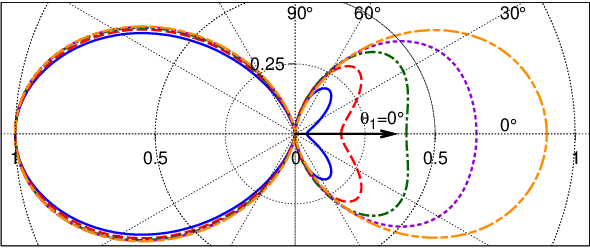}
  \caption{Conditional angular distributions $P(\theta_{12},\theta_1\!=\!0^\circ)$ 
  of the ejected electrons
  for different pulse lengths at $\hw\!=\!70\ev$. The innermost
  (solid blue) line is for $\Tp\!=\!75\as$ FWHM, with
  successive lines for $\Tp\!=\!150\as$, $300\as$, $750\as$, and $4500\as$ 
  FWHM. For better comparison the distributions are normalized
  to a maximum value of one.}
  \label{fig:angdis}
\end{figure}

\begin{figure}[tb]
  \centering
  \includegraphics[width=\linewidth]{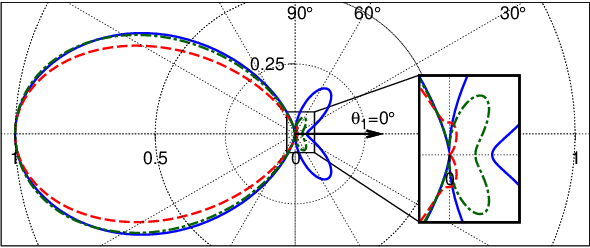}
  \caption{Conditional angular distributions $P(\theta_{12},\theta_1\!=\!0^\circ)$
  for a $75\as$ (FWHM) pulse at $\hw\!=\!70\ev$ (solid blue) and
  for $2\fs$ (FWHM) pulses at $42\ev$ (dashed red) and $52\ev$ (dash-dotted
  green). The distribution for the ultrashort pulse strongly resembles
  the long-pulse distribution in the nonsequential regime ($\hw<54.4\ev$).}
  \label{fig:angdis_seqvsnonseq}
\end{figure}

The attosecond-pulse induced dynamical electron correlation becomes
more clearly visible in the joint angular distribution
$P(\theta_{12},\theta_1)$ (\autoref{fig:angdis}), where $\theta_1$ is
the polar emission angle of one electron with respect to the polarization axis 
of the XUV pulse, $\theta_{12}$ is the angle between the two
electrons, and the energies $E_1,E_2$ are integrated over. 
Here and in the following we choose coplanar geometry with $\phi_1\!=\!\phi_2\!=\!0^\circ$.
In the limit of ``long'' pulses $(\Tp\!=\!4.5\fs$), the joint angular
distribution approaches the product of two independent Hertz dipoles, each of
which signifies the independent interaction of one electron with one
photon. Consequently, also the conditional angular distribution
$P(\theta_{12},\theta_1\!=\!0^\circ)$ corresponds to a Hertz dipole. With
decreasing pulse duration, $P(\theta_{12},\theta_1\!=\!0^\circ)$ is strongly
modified and develops a pronounced forward-backward asymmetry. The
conditional probability for the second electron to be emitted in the
same direction as the first is strongly suppressed.
It is worth noting that this strong preference for back-to-back emission 
for $\Tp\leq150\as$ persists after integration over the electron
energies. Nevertheless, approximately equal energy sharing dominates
(\cf\autoref{fig:energy_spectra}). Thus, the dominant break-up mode
induced by an attosecond pulse corresponds to the ``Wannier ridge''
configuration \cite{Wan1953}. The same break-up mode is observed in the
nonsequential TPDI regime ($\hw<54.4\ev$, \cf\autoref{fig:angdis_seqvsnonseq}),
where the electrons need to exchange energy to achieve double ionization. 
Thus, even in long pulses only electrons ionized within a short time 
of each other can be observed.

\begin{figure}[tb]
  \centering
  \includegraphics[width=\linewidth]{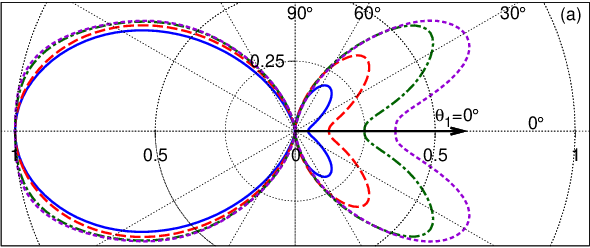}
  \includegraphics[width=\linewidth]{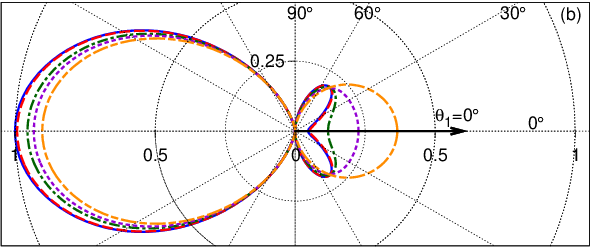}
  \caption{Conditional angular distributions. (a) For a duration (FWHM) of $75\as$ for different photon
  energies. From inside to outside: $70\ev$, $91\ev$, $140\ev$, and
  $200\ev$. The amount of asymmetry decreases with increasing pulse
  energy.
  (b) For different observation times after the $75\as$ FHWM
  pulse at $\hw\!=\!70\ev$. Snapshots were
  taken (from the outermost to the innermost line) immediately at the end of the pulse and
  $50\as$, $150\as$, $600\as$, and $1000\as$ after the end of the pulse.}
  \label{fig:angdis_energies_evolution}
\end{figure}

It is now instructive to inquire into the origin of the strong angular correlations
observed for short pulses. Three different sources can be
distinguished: 

(i) Correlations in the helium ground
state. Due to Coulomb repulsion, the electrons in the ground state
are not independent of each other. For ultrashort pulses,
TPDI can thus be interpreted as a probe 
that maps out the initial-state correlations.

(ii) Induced dipole polarization in the intermediate state. 
When the first electron leaves the core, its electric field
induces polarization of the remaining ion, leading to an asymmetric
probability distribution of the second electron. The second photon then
probes the dynamics in this bound-free complex, such that TPDI can be interpreted
as a pump-probe setup.

(iii) Final-state electron-electron interaction in the
continuum. After the second electron has been released within the
short time interval $\Tp$ as well, their mutual repulsion may redirect
the electrons.

\begin{figure*}[tb]
  \centering
  \resizebox{\linewidth}{!}{
    \includegraphics[height=8cm]{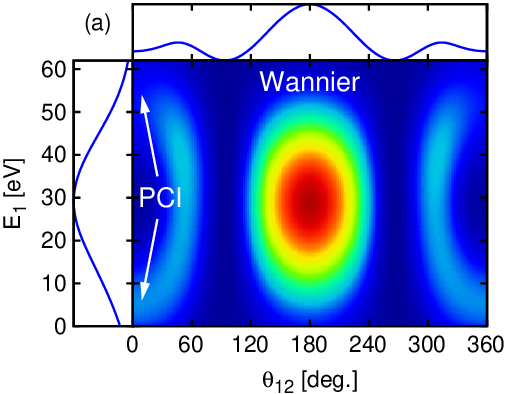}
    \includegraphics[height=8cm]{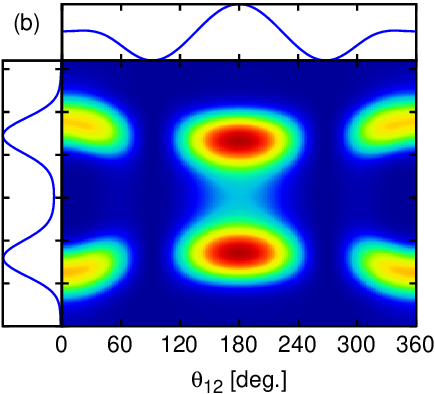}
    \includegraphics[height=8cm]{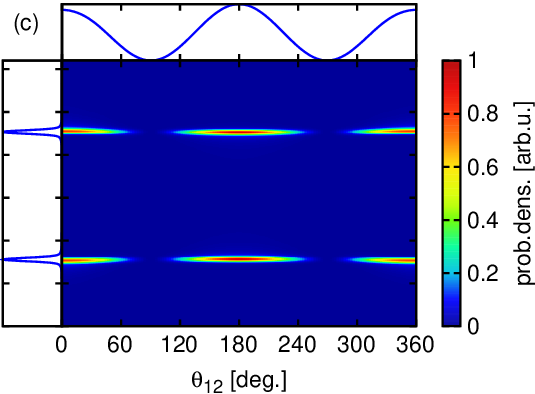}
  }
  \caption{Angle-energy distribution $P(E_1,\theta_{12},\theta_1\!=\!0^\circ)$ in coplanar geometry at
  $70\ev$ photon energy for different pulse durations: (a) $150\as$,
  (b) $450\as$, (c) $4500\as$ FWHM. The side plots
  show the distribution integrated over, respectively, energy and angle.}
  \label{fig:angenedis}
\end{figure*}

While the dividing line between those mechanisms is far from
sharp, the present time-dependent wave packet propagation allows to
shed light on their relative importance since they occur on different
time scales. Relaxation of the ground-state correlations (i) is expected
to occur on the time scale of the orbital period of the residual
electron. As the remaining one-electron wave function will be mostly
in the $n\!=\!1$ and $n\!=\!2$ shells, the timescale for this
relaxation can be estimated as $\ti \!\approx\! \hbar/(E_2 - E_1) \!\approx\! 16\as$,
where $E_n$ is the binding energy in the $n$-th shell of the He$^+$
ion. Therefore, ground-state correlations will become clearly visible
only for pulses with durations shorter than those investigated here.
The time scale for induced dipole polarization (ii) can
be estimated by the time the first electron takes to escape to a
distance where it does not influence the remaining bound electron strongly. Choosing a
somewhat arbitrary distance of $10\au$, the time necessary for the
first electron to reach this distance after absorbing a $70\ev$ photon
is about $120\as$ and thus of the order of the pulse lengths $\Tp$
considered. For higher photon energies, the first electron escapes
more quickly, decreasing the importance 
of this effect. In order to verify this energy dependence, we
have performed calculations at various photon energies for $\Tp\!=\!75\as$.
\autoref{fig:angdis_energies_evolution}a demonstrates that for higher energies
the asymmetry of the joint angular distribution is indeed strongly
reduced. 

Long-range Coulomb interactions in the continuum (iii) extend over
much longer timescales which strongly depend on the relative
emission angles and energies of the electrons, \ie
$\abs{\cvec{k}_1-\cvec{k}_2}$. For example, for two electrons ejected
in the same direction and with similar energies, the interaction will
last much longer than for ejection in opposite directions. This can be
verified by using an ultrashort pulse to start a two-electron wave
packet in the continuum and observing the evolution of the joint
angular distribution after the laser pulse is switched off
(\autoref{fig:angdis_energies_evolution}b). Directly after the pulse, the
distribution of the electrons shows a decreased probability for
ejection on the same side of the nucleus (primarily because of (ii)),
but the lobes in forward and backward direction still mostly retain
the shape expected from a dipole transition. As time passes, continuum final-state
interactions persist and the joint angular distribution develops a
pronounced dip at equal ejection angle. The change at
larger relative angles is almost negligible.

One remarkable feature of the conditional angular distribution is the
persistence of the nodal plane at $\theta\!=\!90^\circ$. While correlation
effects strongly perturb the shape of the independent-particle dipolar
shape, the nodal plane expected for the angular distribution of two
electrons absorbing one photon each is preserved. This is in contrast
to one-photon double ionization, where only one electron
absorbs the photon energy and electron ejection at angles normal
to the polarization axis is indeed observed \cite{BraDoeCoc1998}.
On the other hand, the conditional angular distribution in the nonsequential
regime also exhibits a nodal plane at $\theta\!=\!90^\circ$ 
(\cf\autoref{fig:angdis_seqvsnonseq}).

Additional insights can be gained from a projection of the
two-electron momentum onto the energy-angle plane,
\begin{equation}
  P(E_1,\theta_{12},\theta_1\!=\!0^\circ) = 
  \Int P(E_1,E_2,\Omega_1,\Omega_2) \dE_2 \eqcomma
\end{equation}
in coplanar geometry and for $\theta_1\!=\!0^\circ$. While for long pulses the energy of the emitted
electrons is independent of the relative emission angle
(\autoref{fig:angenedis}c), strong energy-angle
correlations appear for short ($\Tp\leq450\as$) pulses. The dominant
emission channel is the back-to-back emission at equal energy sharing
$(E_1\!\approx\!30\ev)$. This corresponds precisely to the well-known
Wannier ridge riding mode \cite{Wan1953}, previously observed in e-2e
ionization processes \cite{CveRea1974} and also invoked in the
classification of doubly-excited resonances \cite{TanRicRos2000}. Because of
the large instability of the Wannier orbit its presence is more
prevalent in break-up processes than in quasi-bound
resonances. A second subdominant but equally
interesting channel opens for short pulses at $\theta_{12}\!=\!0^\circ$,
\ie emission in the \emph{same} direction. One of the electrons is slowed
down while the other one is accelerated. Hence, the slow electron 
``pushes'' the fast electron from behind, 
transferring part of the energy absorbed from the photon field to the
faster electron. This is the well-known \emph{post-collision
interaction}~\cite{GerMorNie1972,RusMeh1986,ArmTulAbe1987} first
observed by Barker and Berry in the decay of autoionizing states
excited through ion impact~\cite{BarBer1966}.

In conclusion, we have shown that for attosecond XUV pulses the
conventional scenario of ``sequential'' two-photon double ionization
(TPDI) breaks down. Due to the small time interval between the two
photoabsorption processes dynamical electron-electron correlations can
be tuned by the pulse duration $\Tp$. One can view TPDI as an
XUV-XUV \emph{pump-probe} process. The angular and angle-energy
distributions reveal the signatures of electronic correlation induced
by the Coulomb interaction in the intermediate bound-free complex and in the final
state with both electrons in the continuum.
For short pulses, two well-known scenarios, the Wannier ridge riding
mode and the post-collision interaction process, are simultaneously present in the
two-electron emission spectrum.

We thank K.~Ishikawa for interesting discussions. 
J.F., S.N., R.P., E.P., and J.B.\@ acknowledge support by the FWF-Austria, Grant No.\@ SFB016.
Computational time provided under Institutional Computing at Los Alamos.
The Los Alamos National Laboratory is operated by Los Alamos National Security, LLC 
for the National Nuclear Security Administration of the U.S.\@ Department of Energy under Contract No.~DE-AC52-06NA25396.


\end{document}